\documentstyle{article}
\input amssym.def
\input amssym.tex

\def\centreline{\centerline}
\begin{document}
\centreline{\bf{\Large {\em Lsdiff}${\cal M}$ and
the Einstein equations}}
\vskip .5cm
\centreline{\bf{James D.E. Grant}}
\vskip .5cm
\centreline{Department of Physics, University of Newcastle,}
\centreline{Newcastle Upon Tyne, NE1 7RU, United Kingdom.}
\vskip .5cm
\centreline{Departamento de F\'{\i}sica, 
Centro de Investigaci{\'o}n y 
de Estudios Avanzados del I.~P.~N.,} 
\centreline{Apartado Postal~14-740, 
07000 M{\'e}xico D.F., M{\'e}xico.} 
\vskip .5cm
\centreline{Department of Physics and Astronomy, 
University of Pittsburgh, 
Pittsburgh, PA~15260, U.S.A.}
\begin{abstract}
We give a formulation of the vacuum
Einstein equations in terms of a set of
volume-preserving vector fields on a
four-manifold ${\cal M}$. These vectors
satisfy a set of equations which are a
generalisation of the Yang-Mills equations
for a constant connection on flat spacetime.
\end{abstract}

\vskip .5cm

It is known (Mason and Newman~1989) that
the equations which describe self-dual
Ricci-flat metrics can be derived
from the self-dual Yang-Mills
equations in flat space, with a
specific choice of gauge group.
In particular, consider the self-dual
Yang-Mills equations on flat space,
$({\Bbb M}, {\eta})$,
\begin{equation}
F_{ij} = {1 \over 2} {\epsilon}_{ij}{}^{kl}
F_{kl},
\label{SDYM}
\end{equation}
where $F_{ij}$ is the curvature of an
algebra valued connection $A_i$. If we
now impose that the connection $A_i$ is
constant on ${\Bbb M}$, then equations (\ref{SDYM})
become a set of algebraic conditions on the
connection
\[
\left[ A_i, A_j \right] =
{1 \over 2} {\epsilon}_{ij}{}^{kl}
\left[ A_k, A_l \right].
\]
Mason and Newman showed (Mason and Newman~1989) that if we
take the connection $A$ to take
values in {\em Lsdiff}${\cal M}$, the
volume-preserving diffeomorphisms of
an auxiliary four-manifold ${\cal M}$, and\b
write $A_i = ({\bf e}_1, {\bf e}_2,
{\bf e}_3, {\bf e}_4)$, then the
(contravariant) metric ${\bf g} =
{\eta}^{ij} {\bf e}_i \otimes {\bf e}_j$
defines, up to a known conformal factor,
a self-dual metric on the
manifold ${\cal M}$. If we consider
the case where the connection $A$
satisfies the full Yang-Mills equations,
we are led to the equations
\begin{equation}
{\eta}^{ij} \left[ {\bf e}_i,
\left[ {\bf e}_j, {\bf e}_k \right]
\right] = 0,
\label{fullYM}
\end{equation}
which, if we allow the connection 
to have torsion, correspond to Einstein-Cartan
theory (Mason and Newman~1989).

The question we will consider
here is whether it is possible to
find a similar formulation of
the full vacuum Einstein equations.
We begin with four vectors, ${\bf V}_i$,
and an internal metric
${\eta}_{ij}$, with inverse
${\eta}^{ij}$ such that the
(contravariant) metric
\begin{equation}
{\bf g} = {\eta}^{ij}\ {\bf V}_i
\ {\otimes} \ {\bf V}_j
\label{metric}
\end{equation}
satisfies the vacuum Einstein equations. (Letters
$i, j, \dots$ will denote internal indices,
which will be raised and
lowered using ${\eta}^{ij}$ and
${\eta}_{ij}$. We will consider
complex metrics, and will not discuss
reality conditions.)
Defining the structure functions
of the vectors ${\bf V}_i$ by
\[
\left[ {\bf V}_i, {\bf V}_j \right] =
{\tilde C}_{ij}^{\ \ k} {\bf V}_k,
\]
then, by an internal rotation,
we impose the condition that
\[
{\tilde C}_{ij}^{\ \ j} = - {\bf V}_i ( \log f ),
\]
for some function $f$. (This is
possible for {\em all} metrics for some function
$f$. There is no integrability
condition.) We define a
conformally related set of vectors
${\bf e}_{i} = f {\bf V}_i$
with
\[
\left[ {\bf e}_{i}, {\bf e}_{j} \right]
= C_{ij}^{\ \ k} {\bf e}_{k},
\]
where
\[
C_{ij}{}^{k} =
f {\tilde C}_{ij}{}^{k} +
{\bf e}_{i} (\log f)
{\delta}^k_{j} -
{\bf e}_{j} (\log f) {\delta}^k_i.
\]
(The reason for considering this
conformal transformation will
become clear later.) We now note that
\begin{equation}
C_{ij}{}^{j} \equiv {\Gamma}_i =
2 {\bf e}_{i} (\log f).
\label{cijj}
\end{equation}
This means that the ${\bf e}_{i}$
are volume preserving (Mason and Newman~1989).
In particular,
there exists a volume form
${\bf \epsilon} \in {\Lambda}^4({\cal M})$
such that
\[
{\cal L}_{{\bf e}_{i}} {\bf \epsilon} = 0,
\]
where ${\cal L}$ denotes Lie
derivative, and such that
\begin{equation}
{\bf \epsilon}({\bf e}_{1}, {\bf e}_{2},
{\bf e}_{3}, {\bf e}_{4}) = f^2.
\label{FFF}
\end{equation}
(Taking the Lie derivative of equation
(\ref{FFF}) along ${\bf e}_{i}$ then
gives us equation (\ref{cijj}).)
Expressing the Ricci tensor
of the metric (\ref{metric})
in terms of the structure
functions and Ricci tensor of the
vectors ${\bf e}_{i}$, we find that
\begin{eqnarray}
R({\bf V})_{ij}&=&{1\over{f^2}}
\left[R({\bf e})_{ij}+{\eta}_{ij}
\left({\Gamma}^k{\bf e}_{k}(\log f)
-{\eta}^{kl}{\bf e}_{k}{\bf e}_{l}(\log f)
-2{\eta}^{kl}{\bf e}_{k}(\log f)
{\bf e}_{l}(\log f)\right)\right.
\nonumber \\
& &\hskip 15mm\left.
+2{\bf e}_{i}(\log f){\bf e}_{j}(\log f)
+2C^{k}{}_{(ij)}{\bf e}_{k}(\log f)-
2{\bf e}_{(i}{\bf e}_{j)}(\log f)\right].
\label{ricv}
\end{eqnarray}
Using condition (\ref{cijj}),
the equations $R({\bf V})_{ij} = 0$
can be reduced to the form
\begin{equation}
\left[ {\bf e}_{k},
\left[ {\bf e}^k, {\bf e}_{(i} \right] \right]
\cdot {\bf e}_{j)}\ =\
{\eta}_{ij}\ {}^+{\Gamma}_k\ {}^-{\Gamma}^k
\ +\
2\ {}^+ C_{(i}^{\ kl}\ {}^- C_{j)kl}\ -\
2\ {}^+{\Gamma}_{(i}\ {}^-{\Gamma}_{j)},
\label{vacein}
\end{equation}
where we have defined
\begin{eqnarray*}
&{}^{\pm}C_{ij}{}^{k} = {1 \over 2} \left[
C_{ij}{}^{k} \pm
{1 \over 2}{\epsilon}_{ij}{}^{lm}
C_{lm}{}^{k}\right],&
\\
&{}^{\pm}{\Gamma}_i = {}^{\pm}C_{ij}{}^{j},&
\end{eqnarray*}
and
\[
{\bf e}_i \cdot {\bf e}_j \equiv {\eta}_{ij}.
\]
Taking the trace of equation (\ref{vacein}),
we find the condition
for the metric (\ref{metric}) to be scalar flat,
\[
\left[ {\bf e}_i, \left[ {\bf e}^i, {\bf e}_j \right]
\right] \cdot{\bf e}^j = 2 {}^+{\Gamma}_i
{}^-{\Gamma}^i.
\]

\medskip

We thus have:

\medskip

\noindent{\bf Theorem}
{\em
Given a linearly-independent set of
vector fields $\{{\bf e}_i\}$ which
obey equations (\ref{vacein}), and
a volume-form ${\bf \epsilon}$ such
that ${\cal L}_{{\bf e}_i} {\bf \epsilon}
= 0$,
then the set of vector fields
${\bf V}_i = {f}^{-1} {\bf e}_i$
define a vacuum metric, where $f$
is defined by equation (\ref{FFF}).
Conversely, for all vacuum Einstein
metrics, there exists a set of vectors,
$\{{\bf e}_i\}$, (unique up to a restricted
set of internal rotations) with the above
properties.
}

\medskip

This theorem constitutes a generalisation
of the result of Mason and 
Newman (Mason and Newman~1989). They
showed that if the commutator is self-dual,
or anti-self-dual (in the sense that
${}^-C_{ij}{}^{k} = 0$ or
${}^+C_{ij}{}^{k} = 0$, respectively)
then the corresponding metric
(\ref{metric}) is Ricci-flat with
self-dual, or anti-self-dual Weyl tensor,
respectively. In our case, if the commutator
is self-dual or anti-self-dual, then equations
(\ref{vacein}) are satisfied identically
due to the Jacobi identity. In particular,
the right-hand side of equations (\ref{vacein})
is purely an interaction term between the
self-dual and anti-self-dual parts of the
commutator: there is no contribution from
purely self-dual/anti-self-dual fields.
We therefore seem to have
discovered a formalism where the full
gravitational field can be looked on
in terms of an interaction
between its self-dual and anti-self-dual
constituent parts.

The initial motivation for this work
was to see if it was possible to give
a flat space ambi-twistor construction
for the vacuum Einstein equations, in a
spirit similar to the construction for the
full Yang-Mills equations 
(Witten~1978, Green~et~al~1978). The Einstein equations,
as written in (\ref{vacein}), can be looked
on as arising from a generalisation
of the Yang-Mills equations, since
the term on the left-hand side is a
generalisation of the Yang-Mills
operator in flat space for a constant
connection with values in
{\em Lsdiff}${\cal M}$ (c.f. equation
(\ref{fullYM})). (Note that
if we had not performed the
conformal transformation detailed above,
we would be led to an expression which
also involved terms of the form $\left[
{\bf e}_{(i}, \left[ {\bf e}_{j)},
{\bf e}_{k} \right] \right] \cdot {\bf e}^k$,
so the connection with Yang-Mills theory
would be lost.) It is therefore possible
that an ambi-twistor construction for the
vacuum Einstein equations can be found,
via equations (\ref{vacein}), using
similar ideas to those in (Witten~1978, Green~et~al~1978).
This construction would be based upon the
extension of an {\em sdiff}${\cal M}$
bundle over flat ambi-twistor space ${\Bbb A}$,
to a bundle over ${\Bbb P}_3 \times {\Bbb P}_3$,
with the extra condition that the connection
on the diagonal subspace is constant. However,
if we were to simply follow 
(Witten~1978, Green~et~al 1978) and
consider the first obstruction to extending a
given {\em sdiff}${\cal M}$ bundle over
${\Bbb A}$ to a bundle over ${\Bbb P}_3
\times {\Bbb P}_3$, this would lead to a constant
{\em Lsdiff}${\cal M}$-valued connection
which would satisfy the full Yang-Mills
equations on ${\Bbb M}$, as opposed to
equation (\ref{vacein}). We must
split up the four-dimensional vectors ${\bf e}_i$
as a sum of two sets of four eight-dimensional
vectors, so that ${\bf e}_i = {\bf f}_i + {\bf g}_i$ on
${\cal M} \times {\cal M}$. We then wish that equations
(\ref{vacein}) should be satisfied identically on the
diagonal subsurface which defines our original
manifold. There are two main possibilities we
can consider:

\medskip

$\bullet$ We impose that
the commutator of the ${\bf f}_i$ is
self-dual, and the commutator of the
${\bf g}_i$ is anti-self-dual in the sense that
\[
\left[ {\bf f}_i, {\bf f}_j \right] =
{1 \over 2} {\epsilon}_{ij}{}^{kl}
\left[ {\bf f}_k, {\bf f}_l \right], \qquad
\left[ {\bf g}_i, {\bf g}_j \right] =
- {1 \over 2} {\epsilon}_{ij}{}^{kl}
\left[ {\bf g}_k, {\bf g}_l \right].
\]
In order to satisfy equations (\ref{vacein}) on
the diagonal subsurface (Witten~1978, Green~et~al~1978),
however, we would require $\left[ {\bf f}_i,
{\bf g}_j \right] \neq 0$. The value of this
commutator can easily be derived from equations
(\ref{vacein}).

\medskip

$\bullet$ We require that $\left[ {\bf f}_i,
{\bf g}_j \right] = 0$, but impose that the
vectors ${\bf f}_i$ are only gauge equivalent,
under some internal rotation, to a set of
volume-preserving vectors with self-dual
commutator. Similarly, the vectors ${\bf g}_i$
are gauge equivalent to a set of
volume-preserving vectors with
anti-self-dual commutator.

\medskip

The second alternative seems more tempting,
since it maintains the independence of the
self-dual and anti-self-dual parts of the
field. In both the above cases, it remains
to be seen to what extent the resulting
equations can be simplified, or possibly
solved, by a judicious choice of internal
gauges of the two sets of vectors. It seems,
however, that any construction on flat
ambi-twistor space for the vacuum Einstein
equations will be considerably more involved
than that for the Yang-Mills equations. This
would, however, be an alternative to the
deformed ambi-twistor space approach to
the vacuum Einstein equations 
(Isenberg and Yasskin~1982, LeBrun~1985, 
Yasskin~1986, Baston and Mason~1987, Lebrun~1990).

An alternative procedure 
(Chakravarty~et~al~1991, Grant~1993) would be to
choose a suitable coordinate representation
of the vectors ${\bf e}_i$ 
which would lead to a set of differential
equations for various potentials. These equations
would be the generalisation of the Heavenly
equations for half-flat metrics (Pleba{\'n}ski~1975).
It should also be possible to insert the
formalism of Kozameh and Newman 
(Iyer~et~al~1992)
into our equations, which would give a
closed form for (possibly some variant of)
the Light Cone Cut Equation. The formalism
used here suggests it may be worth
investigating a modified version of the
usual spin-coefficient formalism (Newman and Penrose~1962),
where we replace components
of the spin-connection of a
tetrad by the commutator components of our
volume-preserving vectors fields as the
basic objects.

In addition, it may be possible to simplify the
equations of conformal gravity by means
of a well chosen conformal gauge. This
would be a set of fourth order equations
(the vanishing of the Bach tensor (Kozameh~et~al~1985))
for the relevant vector fields, so it is
unlikely that there will be any obvious
relation with the Yang-Mills equations,
although our equations (\ref{vacein}) would,
of course, be a sub-system of these
equations. These, along with the vanishing 
of the Dighton-Eastwood tensor, seem to be the
set of equations that arise naturally
in the deformation of ambi-twistor
spaces (Baston and Mason~1987, 
Lebrun~1990). Alternatively,
from equation (\ref{ricv}), we can
construct the equations for 
certain types of non-Ricci-flat
metrics, e.g. Einstein metrics
with constant non-zero scalar curvature.
It appears, however, that the most
natural description of such metrics
is not in terms of volume-preserving
vectors (Grant~1995).

It should perhaps be pointed out that 
in order to make full contact 
between our equation (\ref{vacein}) and the 
Yang-Mills equations, we would have to 
evaluate the skew-symmetric part of the 
operator %
$\left[ {\bf e}_k, \left[ {\bf e}^k, 
{\bf e}_i \right] \right] \cdot {\bf e}_j$ %
which appears in equation (\ref{vacein}). 
Unfortunately, the skew-symmetric 
part of this operator is not related 
to the Einstein equations (this is 
precisely the reason it does not appear 
in equation (\ref{vacein})). The only 
equations which can be written down 
for the skew-symmetric part, without 
imposing extra constraints on the 
frame, are identities. Therefore, one 
can obtain an expression for the Yang-Mills 
operator, but only at the expense of 
introducing a large amount of redundant 
information. For example, one can show that
\[
\left[ {\bf e}_k, \left[ {\bf e}^k, 
{\bf e}_{[i} \right] \right] 
\cdot {\bf e}_{j]} = 2\ {\bf e}_k 
\left( {}^- C^k{}_{[ij]} \right) 
+ 
{}^-C_{ijk} \ {}^- {\Gamma}^k - 2 
\ {}^-C_{[i|kl|}\  {}^+ C_{j]}{}^{kl}.
\]
This equation is an identity, 
derived using the volume preserving 
condition, and cannot be simplified 
using the Einstein equations. Adding this 
equation to equation (\ref{vacein}) 
and rearranging, we obtain
\begin{eqnarray}
\left[ {\bf e}_k, \left[ {\bf e}^k, {\bf e}_{i} \right] \right] 
&=& {}^+ {\Gamma}_k \ {}^- {\Gamma}^k \ {\bf e}_i 
+ 2\  {}^+ C_{i}{}^{kl} 
\ {}^- C_{j}{}_{kl}\  {\bf e}^j - 2\ {}^+ {\Gamma}_{(i} \ 
{}^- {\Gamma}_{j)} \ {\bf e}^j 
\nonumber \\
& &+ 2\ {\bf e}_k \left( {}^- C^k{}_{[ij]} \right) {\bf e}^j + 
\ {}^- C_{ijk} \ {}^- {\Gamma}^k \ {\bf e}^j. 
\label{vacein2}
\end{eqnarray}
The left hand side of this equation is, in our context, 
the Yang-Mills operator. However, the 
content of equation (\ref{vacein2}) is 
exactly the same as that of equation 
(\ref{vacein}). Moreover, the right 
hand side of the equation is no longer 
a pure interaction term, which is one 
of the main points of interest of 
equation (\ref{vacein}).

Finally, a set of divergenceless vectors
appear naturally in the recent
spin-${3 \over 2}$ work on the vacuum
Einstein equations (Penrose~1994). It would
be interesting to know if these vectors
are related to our ones.

\bigskip

These issues will be considered in more
detail elsewhere.

\bigskip

The author would like to thank R Capovilla, 
H A Chamblin, L E Morales, 
J F Pleba{\'n}ski and K P Tod for 
suggestions. This work was begun 
while the author was in Pittsburgh, 
supported under NSF grant PHY 89-04035. 
It was continued in Mexico, with funding 
from the Centro de Investigacion y de 
Estudios Avanzados del I.P.N. and 
CONACYT. The final version was 
produced in the University of Newcastle, 
under the Sir Wilfred Hall fellowship. 

\section*{References}

\noindent{}Baston R J and Mason L J 1987 
{\it Class. Quantum Grav.} {\bf 4} 815--26

\smallskip

\noindent{}Chakravarty S, Mason L J and Newman E T 1991 
{\it J. Math. Phys.} {\bf 32} 1458--64

\smallskip

\noindent{}Grant J D E 1993 
{\it Phys. Rev.} D{\bf 47} 2606--12

\smallskip

\noindent{}Grant J D E 1995 {\em Some properties of anti-self-dual
conformal structures}.

\smallskip

\noindent{}Green P S, Isenberg J and Yasskin P B 
1978 {\it Phys. Lett.} {\bf 78}B 462--4

\smallskip

\noindent{}Isenberg J A and Yasskin P B 1982
{\it Gen. Rel. Grav.} {\bf 14} 621--7

\smallskip

\noindent{}Iyer S, Kozameh C and Newman E T 1992 
{\it J. Geom. Phys.} {\bf 8} 195--209

\smallskip

\noindent{}Kozameh C, Newman E T and Tod K P 1985 
{\it Gen. Rel. Grav.} {\bf 17} 343--52

\smallskip

\noindent{}LeBrun C R 1985 
{\it Class. Quantum Grav.} {\bf 2} 555--63

\smallskip

\noindent{}LeBrun C R 1990 {\em Twistors in Mathematics and
Physics} ed T N Bailey and R J Baston,
London Mathematical Society Lecture Note Series 156
(Cambridge: Cambridge University Press) pp~71--86

\smallskip

\noindent{}Mason L J and Newman E T 1989 
{\it Comm. Math. Phys.} {\bf 121} 659--68

\smallskip

\noindent{}Newman E T and Penrose R 1962 
{\it J. Math. Phys.} {\bf 3} 566--578

\smallskip

\noindent{}Penrose R 1994 {\it Twistor Newsletter} {\bf 38} 1--9

\smallskip

\noindent{}Pleba{\'n}ski J F 1975 
{\it J. Math. Phys.} {\bf 16} 2395--402

\smallskip

\noindent{}Witten E 1978 {\it Phys. Lett.} {\bf 77}B 394--8

\smallskip

\noindent{}Yasskin P B 1986 {\em
Gravitation and Geometry} ed W Rindler and
A Trautman (Naples: Bibliopolis) pp 477--95

\end{document}